\documentclass[twocolumn]{aastex631}

\usepackage{amsmath}
\usepackage{ulem}
\usepackage{xcolor}
\usepackage{amsmath}
\usepackage{cases}
\usepackage[T1]{fontenc}
\usepackage{subfigure}
\usepackage{epsfig}

\shorttitle{Signature of strange star as the central engine of GRB 240529A}
\shortauthors{Tian et al.}

\begin{document}
\title{Signature of strange star as the central engine of GRB 240529A}

\author{Xiao Tian}
\affiliation{Guangxi Key Laboratory for Relativistic Astrophysics, Department of Physics, Guangxi University, Nanning 530004, China; lhj@gxu.edu.cn}

\author[0000-0001-6396-9386]{HouJun L\"{u}} 
\affiliation{Guangxi Key Laboratory for Relativistic Astrophysics, Department of Physics, Guangxi University, Nanning 530004, China; lhj@gxu.edu.cn}

\author{WenJun Tan} 
\affiliation{Key Laboratory of Particle Astrophysics, Institute of High Energy Physics, Chinese Academy of Sciences, Beijing 100049, China}

\author{ShaoLin Xiong} 
\affiliation{Key Laboratory of Particle Astrophysics, Institute of High Energy Physics, Chinese Academy of Sciences, Beijing 100049, China}

\author{HaoYu Yuan} 
\affiliation{Department of Astronomy, School of Physics, Huazhong University of Science and Technology, Wuhan, 430074, China}

\author{WenYuan Yu}
\affiliation{Guangxi Key Laboratory for Relativistic Astrophysics, Department of Physics, Guangxi University, Nanning 530004, China; lhj@gxu.edu.cn}

\author[0000-0002-1766-6947]{ShuQing Zhong}
\affiliation{School of Science, Guangxi University of Science and Technology, Liuzhou 545006, China}

\author{WenLong Zhang} 
\affiliation{Key Laboratory of Particle Astrophysics, Institute of High Energy Physics, Chinese Academy of Sciences, Beijing 100049, China}
\affiliation{School of Physics and Physical Engineering, Qufu Normal University, Qufu 273165, China}

\author[0000-0002-7044-733X]{EnWei Liang}
\affiliation{Guangxi Key Laboratory for Relativistic Astrophysics, Department of Physics, Guangxi University, Nanning 530004, China; lhj@gxu.edu.cn}

\begin{abstract}
GRB 240529A is a long-duration gamma-ray burst (GRB) whose light curve of prompt emission is composed of a triple-episode structure, separated by quiescent gaps of tens to hundreds of seconds. More interestingly, its X-ray light curve of afterglow exhibits two-plateau emissions, namely, an internal plateau emission that is smoothly connected with a $\sim t^{-0.1}$ segment and followed by a $\sim t^{-2}$ power-law decay. The three episodes in the prompt emission, together with two plateau emissions in X-ray, are unique in the Swift era. They are very difficult to explain with the standard internal/external shock model by invoking a black hole central engine. However, it could be consistent with the prediction of a supramassive magnetar as the central engine, the physical process of phase transition from magnetar to strange star, as well as the cooling and spin-down of the strange star. In this paper, we propose that the first- and second-episode emissions in the prompt $\gamma-$ray of GRB 240529A are from the jet emission of a massive star collapsing into a supramassive magnetar and the re-activity of central engine, respectively. Then, the third-episode emission of prompt is attributed to the phase transition from a magnetar to a strange star. Finally, the first- and second-plateau emissions of the X-ray afterglow are powered by the cooling and spin-down of the strange star, respectively. The observational data of each component of GRB 240529A are roughly coincident with the estimations of the above physical picture.

\end{abstract}

\keywords{Gamma-ray bursts}

\section{Introduction}
The central engine and energy extraction mechanism in gamma-ray bursts (GRBs) study remain an open question \citep{2011CRPhy..12..206Z}. The burst duration ($T_{90}$) of the prompt emission of GRBs ranges from ten milliseconds to several hours, together with the measured redshift, one can roughly estimate the total isotropic energy which ranges from ($10^{\rm 48}-10^{\rm 55}$) erg \citep{2011ApJ...732...29C, 2015PhR...561....1K}. Based on the duration of burst, GRBs can be roughly classified into long-duration GRBs (LGRBs, $T_{90} > 2s$) and short-duration GRBs (SGRBs, $T_{90} < 2s$) \citep{1993ApJ...413L.101K, 2013ApJ...764..179B}. Some LGRBs associated with core-collapse supernovae confirmed that at least some LGRBs originate from the collapse of massive stars (\citealt{1998Natur.395..670G, 2003ApJ...591L..17S, 2004ApJ...609L...5M, 2006ApJ...645L..21M, 2006Natur.442.1011P, 2006Natur.444.1010Z}; \citealt{2008ApJ...689..358R, 2008ApJ...687.1201K}; \citealt{2010ApJ...725.1965L}; \citealt{2012ApJ...759..107K}; \citealt{2019LRR....23....1M}), while the merger of binary compact stars may be the progenitor of SGRBs \citep{2006Natur.444.1010Z, 2010ApJ...725.1965L, 2013ApJ...774L..23B, 2016NatCo...712898J, 2017PhRvL.119p1101A, 2017ApJ...835..181L, 2019ApJ...870L..15L, 2019LRR....23....1M}. Whether it is from a massive star collapse or the merger of binary compact stars, a hyper-accreting black hole \citep{1999ApJ...518..356P, 2013ApJ...765..125L, 2017NewAR..79....1L} or a rapidly spinning, strongly magnetized neutron star (millisecond magnetar)  \citep{1992Natur.357..472U, 1994MNRAS.270..480T, Dai1998a, Dai1998b, 2001ApJ...552L..35Z, 2011MNRAS.413.2031M, 2012MNRAS.419.1537B, 2014ApJ...785...74L, 2015ApJ...805...89L} as the central engine of GRB, may survive after such catastrophic events to drive an extremely relativistic jet and produce GRB.  

On the other hand, a strange star (SS) with an extremely high density has been proposed as the engine of GRBs by many authors \citep{1987PhLB..192...71O, 2000ApJ...541L..71B, 2003ApJ...586.1250B, 2005ApJ...632.1001O, 2016PhRvD..94h3010L, 2016EPJA...52...41D, 2020RAA....20...27O}. Even in the absence of compelling evidence for the existence of strange stars in the universe, there is also no compelling reason why they should not exist. Moreover, the phase transition from neutron star (NS) to strange star (SS) (hereafter NS $\rightarrow$ SS phase transition) is also proposed as a potential energy source for GRBs \citep{1996PhRvL..77.1210C, Dai1998b, 2000ApJ...541L..71B, 2002A&A...387..725O, 2006NCimB.121.1349H}. \cite{1984PhRvD..30..272W} pointed out that strange quark matter (SQM) is composed of nearly equal proportion of $u$, $d$, and $s$ quarks which are more stable than hadronic matter \citep{2005PrPNP..54..193W, 1984PhRvD..30.2379F}. Different mechanisms of phase transition from NS $\rightarrow$ SS have been proposed \citep{1986ApJ...310..261A, 1987PhLB..192...71O, 1996PhRvL..77.1210C, 2004ApJ...614..314B, 2010PhRvD..81l3012M, 2011PhRvD..84h3002H, 2016MNRAS.462L..26P, 2020RAA....20...27O}, but all of them are based on a "seed" of SQM formed inside of a neutron star. The seeds of SQM can be produced via two mechanisms: one is from the early time after the Big Bang; the other is that the seeds are formed in the core of the neutron star due to the NS $\rightarrow$ SS phase transition. The central density of a neutron star continuously increases via the accretion material to increase gravity, or loss the rotation energy of neutron star to decrease its centrifugal force. Once its central density reaches the critical density value of the quantum chromodynamics (QCD) phase transition, the NS will transition to a SS. This deconfinement phase transition lasts only a few milliseconds, but releases a total energy as high as $E_{\rm conv} = (1-4) \times 10^{53}$ erg, which is consistent with the observed $\gamma$-ray energy of GRBs \citep{2000ApJ...541L..71B}.

From the observational point of view, external plateaus with normal decay phases were later commonly observed in Swift early XRT light curves for both LGRBs and SGRBs \citep{2014ApJ...785...74L, 2015ApJ...805...89L}. It can be interpreted as the energy injection into the external shock when a relativistic jet propagates into the circumburst medium \citep{2006ApJ...642..354Z}. However, a good fraction of Swift GRBs exhibit an X-ray plateau followed by a very sharp drop with a temporal decay slope of steeper than 3, called an internal plateau \citep{2007ApJ...665..599T, 2010MNRAS.402..705L, 2013MNRAS.430.1061R, 2017ApJ...835..181L}. Such rapid decay cannot be accommodated by any external shock model, but is consistent with a supramassive magnetar collapsing into a black hole \citep{2018pgrb.book.....Z}, or energy injection from the latent heat released by the solidification of the newborn strange quark star into the GRB afterglow \citep{2011SCPMA..54.1541D,2018ApJ...854..104H}.
 
Recently, a particular long-duration GRB 240529A, was detected by Swift \citep{2024GCN.36556....1E}, Hard X-ray Modulation Telescope (\textit{Insight}-HXMT) \citep{2024GCN.36578....1T}, as well as the Konus-Wind \citep{2024GCN.36584....1S}. The prompt emission of GRB 240529A consists of a triple-episode structure with two long-duration gaps which are quite similar to that of precursors observed in both Burst And Transient Source Experiment (BATSE) \citep{1995ApJ...452..145K} and Fermi/GBM \citep{2018ApJ...862..155L,2020PhRvD.102j3014C}, while the X-ray afterglow light curve exhibits one steep decay and two plateau emissions. The characteristic of two plateau emissions is difficult to explain by existing internal dissipation or external shock models. In this paper, we propose a supramassive magnetar as the central engine. Then, we invoke the physical process of magnetar collapse into strange star, and the cooling and spin-down of the strange star to interpret both the prompt emission and X-ray afterglow of GRB 240529A. The systematic analysis of the observational data for GRB 240529A is shown in Section 2. In Section 3, we attempt to present a physical interpretation of each observed segment of GRB 240529A. The conclusions with some discussion are drawn in Section 4. Throughout the paper, we adopt the convention $Q = 10^{x}Q_{x}$ in cgs units, and employ a concordance cosmology with parameters $\Omega_{\rm M} = 0.30$, $\Omega_{\rm \Lambda} = 0.70$ and $H_{0} = 70 ~\rm km ~\rm s^{-1} ~Mpc^{-1}$. 

\section{Data Reduction and Analysis}
\subsection{Swift/BAT Observations}
At 02:58:31 UT on 29 May 2024 (as the $T_0$), the Swift Burst Alert Telescope (BAT) triggered and located GRB 240529A \citep{2024GCN.36556....1E}. We
downloaded the BAT data from the Swift website\footnote{https://www.swift.ac.uk/archive/selectseq.php?source=obs\&tid=957271} and used the standard HEASOFT tools (version 6.28) to process the BAT data. The BAT light curves in different energy bands are extracted with fixed 1 s time-bin. The light curve shows a single-episode structure with multiple pulses (see Figure 1), and the duration can be estimated as $T_{\rm 90,BAT}=161 \pm 15$ s in the 15-350 keV \citep{2024GCN.36566....1M}. The background is extracted using two time intervals before and after the burst, and then model the background as Poisson noise which is the standard background model for prompt emission in BAT events. More details can refer to \cite{2008ApJS..175..179S}. The time-averaged spectrum of the prompt emission observed by BAT is best fit by a simple power-law model with an index $\Gamma=1.68\pm 0.04$ \citep{2024GCN.36566....1M} due to its narrow energy band.

\subsection{Insight-HXMT Observations}
Hard X-ray Modulation Telescope (\textit{Insight}-HXMT) is the first X-ray astronomical satellite of China, and was successfully launched on June 15, 2017 \citep{2020SCPMA..6349502Z}. It has a very wide energy band from 1 keV to 3 MeV, and includes three kinds of main scientific payloads, such as High Energy X-ray telescope (HE), the Medium Energy X-ray telescope (ME), and the Low Energy X-ray telescope (LE) \citep{2020SCPMA..6349502Z}.

At 02:51:44 UT on 29 May 2024 (corresponding to $T_0-407$ s), \textit{Insight}-HXMT/HE detected GRB 240529A in a routine search of the data \citep{2024GCN.36578....1T}. It is earlier than the triggered time of Swift/BAT about 407 s. The \textit{Insight}-HXMT/HE instrument is equipped with 18 cylindrical NaI(Tl)/CsI(Na) phoswich detectors with a total detection area of 5000 cm$^{2}$. The NaI detectors are sensitive to hard X-rays in the 20–250 keV range, while the CsI detectors serve as anti-coincidence detectors to mitigate upward background noise. Gamma-ray photons with energies exceeding 200 keV can penetrate the spacecraft and payload structure, leaving their signatures in the \textit{Insight}-HXMT/HE system. Due to limitations in crystal thickness and the narrow field of view obstructed by collimators, detecting GRBs with NaI detectors is difficult. However, the CsI(Na) crystals, with their greater thickness, can record high energy gamma-ray photons, thus one can detect the GRBs which are not occulted by the CsI detectors \citep{2020JHEAp..27....1L}.

Then, we extract the light curve of prompt emission of GRB 240529A for all 18 CsI detectors in the 70-900 keV energy band with a time-bin of 1 second. The light curve consists of a triple-episode structure with multiple pulses at $T_0-422$ s, $T_0-315$ s, and $T_0-67$ s, respectively (see Figure 1). The structure of light curve is consistent with that of observations by Konus-Wind \citep{2024GCN.36584....1S}. The duration of the those three episodes (marked as I, II, and III) with an energy range of (150-600) keV are $T_{\rm 90,I}\sim 35$ s, $T_{\rm 90,II}\sim 18$ s, and $T_{\rm 90,III}\sim 180$ s, respectively. 

Finally, we fit the data with a second-order polynomial to obtain a smooth background-subtracted light curve. For each episode ($T_0$-422 s to $T_0$-387 s, $T_0$-315 s to $T_0$-297 s, $T_0$-67 s to $T_0$+113 s), the spectra was extracted using all 18 detectors in the energy range of 150-600 keV, corresponding to the optimal effective area of the CsI detectors. The background was determined by interpolating the signal in two time intervals before and after the burst using a third-order polynomial. More details can referee to \cite{2020JHEAp..27....1L}. The time-averaged spectrum of the first two episodes are best fit by a simple power-law model which is expressed as
\begin{equation}\label{eq:1}
N_{\textrm{PL}}(E)=A\cdot (\frac{E}{100~\mathrm{keV}}) ^{\Gamma}.
\end{equation}
The power-law indices for the first two episodes are 
$\Gamma_{\rm I}=-1.59\pm 0.03$ and $\Gamma_{\rm II}=-1.44\pm 0.09$, respectively. 
The time-averaged spectrum of the third episode, together with Swift/BAT data, can be adequately fitted by a cutoff power-law model,
\begin{eqnarray}\label{eq:2}
N_{\rm CPL}(E) = A\cdot (\frac{E}{100~\mathrm{keV}})^{\Gamma} \mathrm{exp}[-\frac{E(2+\Gamma)}{E_{\rm p}}],
\end{eqnarray}
and one can obtain the $\Gamma_{\rm III}=-1.38\pm 0.04$ and $E_{\rm p}=584^{+102}_{-74}$ keV. The fluence of those three episodes in 10-1000 keV are corresponding to $f_{\rm I}=(1.12\pm 0.09)\times 10^{-5}\rm~ erg/cm^{2}$, $f_{\rm II}=(1.97\pm 0.55)\times 10^{-6}\rm~ erg/cm^{2}$, and $f_{\rm III}=(3.00\pm 0.04)\times 10^{-5}\rm~ erg/cm^{2}$, respectively.

\subsection{Swift X-Ray Telescope Observations}
We downloaded the X-ray telescope (XRT) data of GRB 240529A from the Swift archive\footnote{https://www.swift.ac.uk/xrt\_curves/00998907/}, and the XRT began observing the field at 107 s after the BAT trigger \citep{2024GCN.36564....1D}. Figure 2 shows the XRT light curve in the energy range from 0.3-10 keV. It is worth nothing that the X-ray light curve of GRB 240529A seems to be interesting, and it is quite different from the canonical X-ray emission that is composed of four successive segments \citep{2006ApJ...642..354Z, 2006ApJ...642..389N}. It is composed of two plateau-emission phases and one power-law segment. We perform an empirical fitting to the light curve with one steep decay of power-law and two smooth broken power-law models, which can be expressed as \citep{2007ApJ...670..565L},
\begin{equation}\label{eq:3}
F_1=F_{0,1}\left(\frac{t}{t_{1}}\right)^{-\alpha_1},
\end{equation}
\begin{equation}\label{eq:4}
F_2=F_{0,2}\left[\left(\frac{t}{t_{\rm
b,2}}\right)^{\omega_2\alpha_2}+\left(\frac{t}{t_{\rm
b,2}}\right)^{\omega_2\alpha_3}\right]^{-1/\omega_2},
\end{equation}
\begin{equation}\label{eq:5}
F_3=F_{0,3}\left[\left(\frac{t}{t_{\rm
b,3}}\right)^{\omega_3\alpha_4}+\left(\frac{t}{t_{\rm b,3}}\right)^{\omega_3\alpha_5}\right]^{-1/\omega_3}.
\end{equation}
The total X-ray light curve can be fitting by the superposition of those three models,
\begin{equation}\label{eq:6}
F = F_1 + F_2 + F_3.
\end{equation}
Here, $t_1$ is the starting time of X-ray observations, $t_{\rm b,2}$ and $t_{\rm b,3}$ are the two break times for first plateau and second plateau emissions. $\alpha_1$ is the initial power-law decay index, $\alpha_2$ and $\alpha_3$ are the decay index before and after $t_{\rm b,2}$, and $\alpha_4$ and $\alpha_5$ are the decay index before and after $t_{\rm b,3}$. $\omega_2$ and $\omega_3$ describe the sharpness of the break at
$t_{\rm b,2}$ and $t_{\rm b,3}$, and fixed with $\omega_2=10$ and $\omega_3=7$.

Then, we adopt a Markov Chain Monte Carlo (MCMC) method to fit the light curve by invoking above one power-law $+$ two smooth broken power-law models. The fitting results of X-ray light curve for GRB 240529A are shown in Figure 2, as well as the corner plots of constrained parameters. The first segment is a power-law decay with the temporal index $\alpha_1 = 9.09 \pm 1.45$, which is consistent with the tail emission of prompt $\gamma-$ray from curvature effect \citep{2000ApJ...541L..51K,2007ApJ...666.1002Z, 2015ApJ...808...33U}. Two plateaus primarily dominate the XRT light curve of GRB 240529A. The first plateau exhibits a shallow decay phase followed by a rapid decay phase with the temporal index $\alpha_2 = 0.25 \pm 0.18$ and $\alpha_3 = 3.89 \pm 0.15$, and the break time is $t_{\rm b,2} = 254 \pm 5$ s. In the second plateau, the temporal decay index before and after the break time $t_{\rm b,3} = 14703 \pm 401$ s are $\alpha_4 = 0.09 \pm 0.04$ and $\alpha_5 = 1.95 \pm 0.03$, respectively. The peak flux at the break time of the two plateaus are $F_{\rm 
b,2} = (3.02 \pm 0.14)\times10^{-9} \rm~ erg~cm^{-2}~s^{-1}$ and $F_{\rm 
b,3} = (6.71 \pm 0.30)\times10^{-11} \rm~ erg~cm^{-2}~s^{-1}$, respectively. This fitting parameters are presented in Table 1. 

\subsection{Optical follow-up Observations}
Since the Swift/BAT triggered GRB 240529A, a number of optical telescopes are follow-up to observe, such as the Gravitational-wave Optical Transient Observer (GOTO; \citealt{2024GCN.36559....1K}), SAO RAS 1m telescope \citep{2024GCN.36601....1M}, AZT-33IK telescope \citep{2024GCN.36585....1P}, 1.3m Devasthal Fast Optical Telescope (DFOT; \citealt{2024GCN.36589....1R}), Skynet's 0.6m RRRT telescope \citep{2024GCN.36568....1D}, 50cm-B, 50cm-C, 100cm-C telescopes \citep{2024GCN.36734....1A}. However, most optical telescopes follow-up observations only present one more observed point or upper limits. So that, we collect the observations of $R$-band which is the most observed band from the GCN reports, and the plot the optical light curve.

\subsection{Redshift measured of GRB 240529A}
The spectroscopy of the afterglow of GRB 240529A is obtained by using OSIRIS at the 10.4 m GTC telescope in the Roque de los Muchachos Observatory, and the spectrum shows a bright continuum with many strong spectral features\citep{2024GCN.36574....1D}. Based on the observed three absorption lines, they identified the redshift of GRB 240529A as $z=2.695$, $z=2.035$, and $z=1.695$, respectively \citep{2024GCN.36574....1D}.

By adopting the redshift of GRB 240529A as $z=2.695$, $z=2.035$, and $z=1.695$, together with the fluence of three-episode emissions in the prompt emission and fitting results in X-ray afterglow emission, one can calculate the isotropic energy of three-episode emissions and two plateau emissions (see Table 2), 
\begin{equation}\label{eq:7}
E_{\rm \gamma,iso,i} = 4 \pi D^{2}_{\rm L} f_{\rm i} / (1+z),
\end{equation}
where $i=$I, II, or III. We correct the isotropic energy of three-episode emissions to the beaming-correction values $E_{\rm \gamma}$ by considering a possible beaming correction factor: 
\begin{equation}\label{eq:8}
f_{\rm b} = 1 - cos\theta_{\rm j} \simeq (1/2) \theta^2_{\rm j},
\end{equation}
where $\theta_{\rm j}$ is the jet opening angle, and $E_{\rm \gamma} = E_{\rm \gamma,iso} f_{\rm b}$.

The isotropic break luminosity at the break time of two plateau emissions can be calculated by 
\begin{equation}\label{eq:9}
L_{\rm b,iso, j} = 4 \pi D^{2}_{\rm L} F_{\rm b, j} / (1+z),
\end{equation}
where $j=$1 or 2 indicate the first and second plateau emission, respectively. Then, the isotropic energy of X-ray plateau emission, $E_{\rm X,iso, j}$, can be derived by employing the break time and break luminosity \citep{2014ApJ...785...74L},
\begin{equation}\label{eq:10}
E_{\rm X,iso,j} \simeq L_{\rm b,iso,j} \cdot \frac{t_{\rm b,j}}{(1+z)}.
\end{equation}

One needs to note that the first episode emission is occurred before the BAT trigger time about 407 s. So that, the true break time of two plateau emissions should be $t_{\rm b, j} + 407$ s. If this is the case, one can estimate the isotropic energy of two plateau emissions, and the calculated results for different redshift of GRB 240529A are also shown in Table 2.

\section{Physical Interpretation}
In this section, we invoke the supramassive magnetar as the central engine, phase transition from magnetar to strange star, the cooling and spin-down of the strange star to interpret each observed segment of prompt emission and X-ray afterglow of GRB 240529A. 

\subsection{The first and second episode of prompt emission: jet radiation of central engine}
The long-duration prompt $\gamma-$ray of GRB 240529A suggests that it is likely related to the deaths of massive stars. If this is the case, a hyper-accreting black hole, a rapidly spinning magnetar, or a strange star as the central engine may be formed \citep{2018pgrb.book.....Z}. The central engine lasts long enough to allow the jet head to break out of the star, and a successful jet is produced. Based on the feature of X-ray afterglow discussed below, the supramssive magnetar with surface polar cap magnetic field $B_{\rm p}$ of about $10^{15}$ G and initial spin period $P_0$ of about 1 ms is a potential candidate as the central engine of GRB 240529A. The newborn millisecond magnetar will power the GRB jet by losing its huge total rotational energy $E_{\rm rot}$. 

Based on the method in \cite{2001ApJ...552L..35Z}, the total rotational energy of the newborn millisecond magnetar can be expressed as:
\begin{equation}\label{eq:11}
E_{\rm rot} = \frac{1}{2} I \Omega_{0}^2 \approx 2 \times 10^{52}~\rm erg~ M_{1.4} R^2_{6} P^{-2}_{0,-3},
\end{equation}
where I is the moment of inertia; and $\Omega_{0}$, $R = 10~\rm km$ and  $M_{1.4} = M/1.4 M_{\rm \odot}$ are the initial angular frequency, radius and mass of the neutron star, respectively. There are three energy extraction mechanisms for a millisecond magnetar engine to power the GRB jet, such as rotation energy \citep{1983JBAA...93R.276S, 2001ApJ...552L..35Z}, magnetic bubble eruption due to differential rotation \citep{1998ApJ...505L.113K, 2000ApJ...542..243R}, and accretion of neutron star \citep{2008ApJ...683..329Z, 2009ApJ...703..461Z}.

The isotropic energy $E_{\rm \gamma,iso, I}$ of the first episode of GRB 240529A prompt emission is about $2\times10^{53}$ erg, $10^{53}$ erg, and $8\times10^{52}$ erg at redshift $z=2.695$, $z=2.035$, and $z=1.695$, respectively. Since the $\theta_{\rm j}$ of GRB 240529A is unknown, we have to adopt a typical value of the beaming factor as $f_{\rm b} = 0.01$ to do the jet correction of $E_{\rm \gamma, I}$  \citep{2008ApJ...675..528L, 2009ApJ...698...43R, 2012Sci...338.1445N,2014ApJ...785...74L, 2017ApJ...835..181L}. The values of $E_{\rm \gamma, I}$ for different redshifts $z=2.695$, 2.035, and 1.695 can be estimated as $2\times10^{51}$ erg, $10^{51}$ erg, and $8\times10^{50}$ erg, respectively. On the one hand, the total energy budget of the newborn magnetar can be satisfied with the observed energy released in the first episode of prompt $\gamma-$ray emission. On the other hand, together with considering the feature of two X-ray plateau emissions discussed below, it suggests that the first episode of prompt emission is possibly produced by the supramassive magnetar central engine from the death of massive star.

The intensity of second episode emission is five times less than that of first episode emission, and this feature is not surprising in prompt emission of GRBs \citep{1995ApJ...452..145K, 2005MNRAS.357..722L, 2013ApJ...775...67B, 2014ApJ...789..145H, 2018ApJ...862..155L}. 
\cite{2018ApJ...862..155L} showed that there are not any significant difference in the spectral and temporal properties of these two sub-burst emissions, and suggested that these two sub-burst emissions likely share the same physical origin. On the other hand, the X-ray flares have been discovered in a good fraction of GRBs afterglow \citep{2005Sci...309.1833B, 2007ApJ...671.1903C, 2010MNRAS.406.2149M}. Temporal and spectral analyses of X-ray flares reveal many properties analogous to prompt emission and suggest that X-ray flares are directly powered by the GRB central engine, similar to prompt emission. They are the extension and delay of prompt emission, and are possible from the later re-activity of the central engine \citep{2005Sci...309.1833B, 2005MNRAS.364L..42F, 2006ApJ...642..354Z}.

If the second-episode emission of GRB 240529A is indeed from the later central engine re-activity. Generally speaking, in order to produce a second “burst” as is observed in GRB 240529A, the central engine must restart about one fifth energy by comparing the first burst. Several models are proposed to interpret such sub-burst in the prompt emission of GRBs, such as fragmentation in the massive star envelope \citep{2005ApJ...630L.113K}, fragmentation in the accretion disk \citep{2006ApJ...636L..29P}, the magnetic barrier around the accretor \citep{2006MNRAS.370L..61P}, fall-back accretion of a newborn millisecond magnetar \citep{2011ApJ...736..108P, 2014MNRAS.438..240G, 2017MNRAS.470.4925G, 2018MNRAS.478.4323G, 2020MNRAS.492.3622L, 2024ApJ...962....6Y}.

However, those models also predict that the X-ray flares should be produced at a late time, but we do not observe any flares in the X-ray afterglow. Two ways may be used to solve the above inconsistency: one is that it indeed produces the X-ray flares sometimes before the magnetar transforms into a strange star, but XRT does not start to observe; The other one is that the produced X-ray flares are too weak to be detected by XRT. In any case, the later central engine re-activity model is not contradictory to the observations of the second episode for GRB 240529A.

\subsection{The third episode of prompt emission: phase transition from magnetar to strange star}
Theoretically, the strange star is with an extremely high density, and a three-flavor ($uds$) quark-gluon plasma is more stable than baryons and a two-flavor ($ud$) QGP. An entire neutron star may be converted to a star made of strange quark matter. The release of a substantial amount of energy during NS $\rightarrow$ SS phase transition to power the GRB prompt emission has been extensively researched as a potential energy source of GRBs \citep{1996PhRvL..77.1210C, Dai1998b, 2002A&A...387..725O}.

Therefore, in this section, we propose that the third-episode emission of GRB 240529A prompt emission is powered by the phase transition from a neutron star to a strange star. The possible physical process can be described as follows: 

(a) The neutron star formed by the gravitational collapse of a massive star results in an increasing density of its center via accretion. Once the central density reaches the critical value for the quark deconfinement phase transition, a seed of SQM will be generated within the core. This process takes about a few milliseconds.

(b) Then, due to the production and diffusion of these seeds of SQM, a neutron star transforms into a strange star after tens of seconds. The surface layer of the newborn SS is very hot with a temperature $\sim 10^{11}$ K. 

(c) The baryonic material can be ablated again from the newborn SS surface, as long as the surface is not completely filled with strange quark matter \citep{2016EPJA...52...41D}. In this way, a new $\gamma-$ray emission can be produced by the strange star. 

Here, we assume that the generated strange star has a mass of $M_{\rm SS}=2.5~\rm M_{\odot}$ (about $5 \times 10^{33}~\rm g$), and it includes the total baryon number $n \sim 2.5\times10^{57}$. If each nucleon can release 50 MeV heat energy during the phase transition, the expected energy can be estimated as
\begin{equation}\label{eq:12}
E_{\rm conv} \sim 50\times2.5\times10^{57} \rm MeV = 2\times10^{53} \rm erg.
\end{equation}
On the other hand, by considering the different equation of state of NS and SS, \cite{2000ApJ...541L..71B} provided that the energy released during the phase transition range of $(1-4) \times 10^{53}$ erg.

Observationally, the calculated isotropic energy of the third-episode emission of prompt $\gamma-$ray $E_{\rm \gamma,iso,III}$, are about $5\times10^{53}$ erg, $3\times10^{53}$ erg, and $2\times10^{53}$ erg at redshift $z=2.695$, $z=2.035$, and $z=1.695$, respectively. By considering the beaming correction factor $f_{\rm b} = 0.01$, one can obtain the beaming-corrected energy $E_{\rm \gamma,III}$ of this episode which is in the range of $(2-5)\times 10^{51}$ erg. It is consistent with that of energy predicted during phase transition for energy conversion efficiency $\epsilon=0.01$. It means that the phase transition at least from the side of energy released could reasonably explain the third-episode emission of GRB 240529A prompt emission. If this is the case, a supramassive strange star as the central engine of GRB 240529A may survive to drive third-episode $\gamma-$ray emission.

\subsection{The first-plateau emission of X-ray afterglow: cooling of strange star}
Initially, a newborn strange star may be in the liquid phase due to its high temperature. The liquid strange star will transition into a solid state when the temperature is decreased due to the energy released by cooling \citep{2003ApJ...596L..59X, 2009ScChG..52..315X}. The latent heat released energy during the cooling is proposed to explain the observed X-ray plateau followed by a rapid decay phase in GRBs (or internal plateau), and this latent heat released is injected into the GRB afterglow through a mechanism which is similar to a Poynting flux-dominated outflow  \citep{2011SCPMA..54.1541D, 2018ApJ...854..104H}.

Based on the solid SS model proposed by \cite{2009MNRAS.398L..31L}, one can calculate the released energy ($E$) by each baryon during the transition from liquid to solid of the strange star. By fixing the depth of the potential $U_{0}=$100 MeV, and the ratio between melting heat and the potential $f_{\rm mp} \approx 0.01-0.1$ \citep{2011SCPMA..54.1541D}, it can be written as
\begin{equation}\label{eq:13}
E \sim f_{\rm mp}U_{0} \approx (1 - 10)~\rm MeV.
\end{equation}
For $M_{\rm SS}=2.5~\rm M_{\odot}$ with $n \sim 2.5\times10^{57}$ of strange star, the total released energy during cooling can be estimated as
\begin{equation}\label{eq:14}
E_{\rm cooling} = nE \approx (0.4-4)\times10^{52}~\rm erg.
\end{equation}

By considering the energy conversion efficiency $\xi=0.2$ from cooling energy to X-ray emission, the radiation energy in X-ray emission should be in the range of $(0.8-8)\times10^{51}~\rm erg$. Moreover, the observed isotropic energy of the first-plateau emission $E_{\rm X,iso,1}$ of X-ray aferglow for GRB 240529A is distributed within the range of $(5-9)\times10^{51}$ erg at three different redshifts ($z=2.695$, 2.035, and 1.695). It is found that the cooling energy released is consistent with the observed X-ray emission of GRB 240529A. 

On the other hand, the radiation timescale for the strange star can be estimated as follows by assuming blackbody radiation:
\begin{equation}\label{eq:15}
t = \frac{E_{\rm cooling}}{L_{\rm rad}}=\frac{E_{\rm cooling}}{\sigma T^{4} 4 \pi R^{2}},
\end{equation}
where $\sigma$ and $T$ are the Stefan–Boltzman constant and temperature of the blackbody, respectively. $R$ is the radius of the newborn SS. More significantly, when the strange star is cooling from liquid to solid phase, the radiation luminosity $L_{\rm rad}$ is almost constant because the temperature is not changed. So it would appear a plateau emission in the light curve of X-ray afterglow. At the end of the cooling, the energy injection will be abruptly cutoff, because the solid SS with low heat capacity is cooling very fast. By adopting the duration of plateau timescale $t=t_{b,1}\sim 661$ s and conversion efficiency $\xi=0.2$, one can estimate the temperature $kT\sim 2$ MeV, 1.85 MeV, and 1.7 MeV at different redshift $z=2.695$, 2.035, and 1.695, respectively. It is natural to explain the presence of a plateau with a rapid decay (first-plateau) phase in the X-ray light curve.

\subsection{The second-plateau emission of X-ray afterglow: magnetic dipole radiation of strange star spin-down}
After the strange star cooling, the newborn strange star has cooled to a solid-crystal. We consider that the properties of the solid SS are similar to that of the magnetar. The strange star with rapid spinning as the GRB central engine can also lose its rotational energy mainly via two mechanisms, i.e., electromagnetic (EM) dipole and gravitational wave (GW) radiations \citep{2001ApJ...552L..35Z, 2013PhRvD..88f7304F, 2016MNRAS.458.1660L}. For EM-dominated, the temporal evolution of the EM luminosity $L_{\rm EM}$ can be formulated as \citep{2016MNRAS.458.1660L, 2018MNRAS.480.4402L}
\begin{equation}\label{eq:16}
L_{\rm EM}=L_{0}(1+t/\tau_{\rm EM})^{-2}.
\end{equation}
For GW-dominated, the temporal evolution of the GW luminosity $L_{\rm GW}$ can be expressed as
\begin{equation}\label{eq:17}
L_{\rm GW}=L_{0}(1+t/\tau_{\rm GW})^{-1}.
\end{equation}
Here, $L_{0}$ is the characteristic spindown luminosity. $\tau_{\rm EM}$ and $\tau_{\rm GW}$ are characteristic spindown time scale for the case of EM-dominated and GW-dominated, respectively.

We note that the decay index of the second-plateau emission is about $L\propto t^{-2}$ based on the fitting in Section 2.3, and it is consistent with the case of energy loss dominated by EM dipole radiation. So that, the $\tau_{\rm EM}$ and $L_{0}$ can be expressed as 
\begin{equation}\label{eq:18}
\tau_{\rm EM} = 2.05 \times 10^{3}~ I_{45}B^{-2}_{p,15}P^{2}_{0,-3}R^{-6}_{6}~\rm s,
\end{equation}
\begin{equation}\label{eq:19}
L_{0} = 1.0 \times 10^{49}~ B^2_{p,15}P^{-4}_{0,-3}R^6_{6}~\rm erg~s^{-1}.
\end{equation}
For EM-dominated, we adopt $L_{\rm EM} \approx L_{0}$, and the isotropic break luminosity $L_{\rm b,iso,2}$ of second plateau emission is about $1.08\times10^{48}$ erg, $0.67\times10^{48}$ erg, and $0.48\times10^{48}$ erg at redshift $z=2.695$, 2.035, and 1.695, respectively. Here, after the jet-correction of break luminosity $L_{\rm b,2}$ can be expressed as $L_{\rm b,2} = L_{\rm b,iso,2}f_{\rm b}$. By considering the energy conversion efficiency ($\eta=0.01$) from strange star spin down to the X-ray emission, the break luminosity of second-plateau emission can be written as $L_{\rm b,2} = \eta L_{\rm 0}$. Based on Eq.(\ref{eq:18}) and Eq.(\ref{eq:19}), by adopting the characteristic spin-down timescale $\tau_{\rm EM} \sim t_{\rm b,2}/(1+z)$ and conversion efficiency $\eta = 0.01$, together with the observed second-plateau luminosity of X-ray afterglow, one can constrain the characteristic parameters of the strange star. In order to compare the difference between the strange star as central engine of GRB 240529A and other candidates of magnetar central engine of LGRBs from \cite{2014ApJ...785...74L}, we similarly adopt the jet opening angle to be $5^{\circ}$, and then do the jet-correction of the derived $B_p$ and $P_0$ for GRB 240529A. One has $B_p=(2.40\pm0.14)\times10^{15}$ G and $P_0=3.40\pm0.09$ ms at redshift $z = 2.695$, $B_p=(2.52\pm0.14)\times10^{15}$ G and $P_0=3.93\pm0.10$ ms at redshift $z = 2.035$, and $B_p=(2.64\pm0.15)\times10^{15}$ G and $P_0=4.36\pm0.11$ ms at redshift $z = 1.695$. Figure 3
shows the distribution of $B_{\rm p}$ and $P_{0}$ of GRB 240529A with different redshifts, and compare with other candidates of magnetar central engine of LGRBs from \cite{2014ApJ...785...74L}. We find that the derived $B_{\rm p}$ and $P_{0}$ fall into the reasonable range at different redshifts, and mix with those of the other LGRBs candidates with magnetar central engine from \cite{2014ApJ...785...74L}.

\section{Conclusion and Discussion}
GRB 240529A is a long-duration GRB that was detected by Swift, \textit{Insight}-HXMT, as well as the Konus-Wind \citep{2024GCN.36556....1E,2024GCN.36578....1T,2024GCN.36584....1S}. The possible redshift of GRB 240529A is identified as $z=2.695$, 2.035, and 1.695, based on three absorption lines in the spectrum \citep{2024GCN.36574....1D}. \textit{Insight}-HXMT/HE detected GRB 240529A in a routine data search; it was earlier than the triggered time of Swift/BAT about 407 s \citep{2024GCN.36578....1T}. The prompt emission of GRB 240529A detected by \textit{Insight}-HXMT is composed of a triple-episode structure, separated by quiescent gaps of tens to hundreds of seconds. Only the third-episode emission is simultaneously observed by both Swift/BAT and \textit{Insight}-HXMT. By extracting the time-averaged spectrum of each episode of the prompt emission of GRB 240529A, we find that a simple power-law model can best fit the first two episodes, while the third episode can be adequately fitted by a cutoff power-law model with $E_{\rm p}=584^{+102}_{-74}$ keV. 

More interestingly, the X-ray afterglow light curve of GRB 240529A exhibits the behavior of a steep decay phase and two plateau phases. We adopt a MCMC method to fit the light curve by invoking one power-law $+$ two smooth broken power-law models. The first segment is a power-law decay with the temporal index $\alpha_1 = 9.09 \pm 1.45$ that is consistent with the tail emission of the prompt $\gamma-$ray due to the curvature effect \citep{2000ApJ...541L..51K,2007ApJ...666.1002Z, 2015ApJ...808...33U}. The first plateau emission with the temporal decay index and the break time are $\alpha_2 = 0.25 \pm 0.18$, $\alpha_3 = 3.89 \pm 0.15$, and $t_{\rm b,2} = 254 \pm 5$ s, which is consistent with the internal plateau. The temporal decay indices of the second plateau before and after the break time $t_{\rm b,3} = 14703 \pm 401$ s are $\alpha_4 = 0.09 \pm 0.04$ and $\alpha_5 = 1.95 \pm 0.03$, respectively. 

Such a feature of three episodes in the prompt emission of GRB 240529A, together with two plateau emissions in the X-ray afterglow, suggests that the standard internal/external shock model of a black hole central engine is very difficult to explain both the prompt emission and X-ray afterglow of GRB 240529A. However, it could be more consistent with the prediction of a supermassive magnetar as a central engine, physical process of magnetar collapse into strange star, and the cooling and spin-down of strange star. 

In this paper, we propose the above physical picture to interpret each observed segment in both the prompt emission and X-ray afterglow of GRB 240529A, and it is summarized as follows:

(1) A supramassive magnetar as the central engine may be formed after the death of a massive star. The central engine lasts long enough to allow the jet head to break out the star, and a successful jet is produced by losing its huge total rotational energy $E_{\rm rot}$. This is naturally to explain the first episode of prompt emission. Then, the central engine is re-activity to power a weak sub-burst via fragmentation in the accretion disk \citep{2006ApJ...636L..29P}, magnetic barrier around the accretor \citep{2006MNRAS.370L..61P}, or fallback of magnetar \citep{2011ApJ...736..108P, 2014MNRAS.438..240G, 2017MNRAS.470.4925G, 2018MNRAS.478.4323G, 2020MNRAS.492.3622L, 2024ApJ...962....6Y}. It can be used to explain the second-episode emission of prompt emission of GRB 240529A.

(2) Theoretically, the strange star is more stable than neutron star, and a substantial amount of energy may be released when a magnetar transitions into a strange star via quark deconfinement phase transition. The third-episode emission of prompt $\gamma-$ray can be powered by the phase transition from a neutron star to a strange star.

(3) Initially, a newborn strange star may be in the liquid phase due to its high temperature. The energy lost via some cooling mechanisms can result in the newborn strange star transforming from liquid to solid state. It predicts a plateau emission followed by an extremely sharp decay, which is consistent with the observations of first-plateau emission of X-ray afterglow.

(4) After the strange star cooling, the newborn strange star has cooled to a solid-crystal. The properties of the solid strange star are similar to those of the magnetar. The strange star with rapid spinning as the GRB central engine can also lose its rotational energy mainly via magnetic dipole radiation. The model predicts that the temporal evolution of the EM luminosity is $L_{\rm EM}=L_{0}(1+t/\tau_{\rm EM})^{-2}$, which is consistent with the observed second-plateau emission of X-ray afterglow.

If this is the case, we also infer the magnetic field strength $B_{\rm p}$ and initial spin period $P_{0}$ of the strange star of GRB 240529A, and compare these values with other candidates of magnetar central engine for LGRBs from \cite{2014ApJ...785...74L}. We find that the derived $B_{\rm p}$ and $P_{0}$ fall into the reasonable range at different redshifts, and mix with those of other LGRBs candidates with magnetar central engine.

If the observed prompt emission and X-ray afterglow are all contributed from the internal dissipation. One question is that where is external component when the ejecta interact with interstellar medium? In order to answer this question, we collect the observed optical data in $R-$band which are the most data points released from GCN report \citep{2024GCN.36559....1K,2024GCN.36601....1M,2024GCN.36585....1P,2024GCN.36589....1R,2024GCN.36568....1D,2024GCN.36734....1A}. The optical light curve shows an initial bump connected to a normal decay, and it seems to be contributed from the external shock model. Therefore, we adopt the standard afterglow model to fit the optical light curves for three different redshifts \citep{1998ApJ...497L..17S, 1999MNRAS.309..513H}, and the fitting results are shown in Figure 4. One can obtain the parameters of external shock model as the initial Lorentz factor $\Gamma = 71$, the circumburst medium density $n=10$, the equipartition parameters $\epsilon_{\rm e}=\epsilon_{\rm B} = 0.08$, and the electron injection spectral index $p = 2.8$. The initial total kinetic energy $E_{\rm K,iso}$ is about $5\times 10^{52}$ erg, $2.5\times10^{52}$ erg, $2.0\times10^{52}$ erg at redshift $z=2.695$, 2.035, and 1.695, respectively. We find that the external component in X-ray afterglow still exist, but it is too faint to be detected by comparing with internal dissipation.

Moreover, the physical explanation of GRB 240529A what we adopted may not be the only one to the observations, but it should be the most self-consistent explanation with the observations. For example, \cite{2024ApJ...976L..20S} invoked two shocks launched from the central engine separately to explain the multi-wavelength emissions of GRB 240529A, but is not to explain the first X-ray plateau emission. Also, \cite{2023ApJ...955...93A} recently proposed to adopt a binary driven hypernova (BdHN) model to explain GRB 171205A whose X-ray light curve (e.g., a two-plateau emission with sparse data) is quite similar to that of GRB 240529A, but is not related to the three-episode emission in prompt emission \citep{2014A&A...565L..10R}. \cite{2007ApJ...670.1247W} proposed the jet bow shock and relativistic jet itself to explain two-episode emission, but is not related to two-plateau X-ray emissions. \cite{2008MNRAS.383.1397L} proposed a simple non-stationary three-parameter collapse model to explain both precursors and X-ray plateau emission, or continues to eject relativistic material from the central engine \citep{2014MNRAS.445.1625N}.

On the other hand, two interesting questions naturally arise. One is why it occurs in GRB 240529A and not in every GRB. For this question, there are two possible ways to interpret it, such as the probabilities of a massive star collapsing into a black hole, as well as produced GRBs are slightly higher than that of collapsing into a magnetar to produce GRB \citep{2022ApJ...938...69P}, or the strict requirements for initial mass of a massive star and the equation of state of a neutron star. Another one is that how to identify the central engine (e.g., neutron star or black hole) of GRBs through the electromagnetic observations is an extremely challenging task. Recently, \cite{2024Sci...383..898F} confirmed that the central engine of SN 1987A is a neutron star with the observed narrow infrared emission lines of argon and sulfur from the James Webb Space Telescope (JWST). Moreover, the GRB 170817A/GW170817 event from the merger of neutron stars is currently the only example of joint electromagnetic and gravitational wave observations to confirm a Kerr black hole as the central engine \citep{2023A&A...669A..36V}. For the long-duration GRBs, we expect to detect both electromagnetic and gravitational wave for the same burst event in the future, and it will provide decisive evidence to answer this question.

\begin{acknowledgements}
We acknowledge the use of the public data from the Swift Science Data Center. This work is supported by the Guangxi Science Foundation (grant No. 2023GXNSFDA026007), the Natural Science Foundation of China (grant Nos. 12494574, 11922301 and 12133003), the Program of Bagui Scholars Program (LHJ), and the Guangxi Talent Program (“Highland of Innovation Talents”). S.Q.Z is supported by the starting Foundation of Guangxi University of Science and Technology (grant No. 24Z17).
\end{acknowledgements}

{}

\clearpage
\begin{table*}[h]\footnotesize %
 \centering
  \caption{Fitting Results of the X-ray Afterglow Light Curves of GRB 240529A}
  \setlength{\tabcolsep}{2mm}{
  \begin{center}
  \renewcommand\arraystretch{1.5}
  \begin{tabular}{c|c|c|ccc}
  \hline
  \hline
            & $\alpha$ &  $t_{\rm b}$ & $F_{\rm b}$ \\
            &          &    (s)       & ($\rm erg~cm^{-2}~s^{-1}$) \\
  \hline
  Power-law &	$\alpha_1 = 9.09 \pm 1.45$ &            & $(8.51 \pm 0.75)\times10^{-9}$\\ 
  \hline
  First plateau & $\alpha_2 = 0.25 \pm 0.18$ & $t_{\rm b,1} = 254 \pm 5$ & $(3.02 \pm 0.14)\times10^{-9}$ \\
              & $\alpha_3 = 3.89 \pm 0.15$ &  &   \\
  \hline
  Second plateau & $\alpha_4 = 0.09 \pm 0.04$ & $t_{\rm b,2} =14703 \pm 401$ & $(6.71 \pm 0.30)\times10^{-11}$ \\
              & $\alpha_5 = 1.95 \pm 0.03$  &      &\\
  \hline
\end{tabular}
\end{center}}
\end{table*}

\begin{table*}[h]\footnotesize %
 \centering
  \caption{Calculation results of the prompt and afterglow emissions of GRB 240529A}
  \setlength{\tabcolsep}{2mm}{
  \begin{center}
  \renewcommand\arraystretch{1.5}
  \begin{tabular}{c|c|c|c|c|c|c|cc}
  \hline
  \hline
     redshift &  I &  II &  III   &     \multicolumn{2}{c}{First plateau}  &  \multicolumn{2}{|c}{Second plateau}\\
   \cline{5-8}
           & $E_{\rm \gamma,iso,I}$ & $E_{\rm \gamma,iso,II}$ & $E_{\rm \gamma,iso,III}$  & $L_{\rm b,iso,1}$ & $E_{\rm X,iso,1}$   & $L_{\rm b,iso,2}$ & $E_{\rm X,iso,2}$ \\
            & ($\rm 10^{52}~erg$) & ($\rm 10^{52}~erg$) & ($\rm 10^{52}~erg$) &  ($\rm 10^{48}~erg~s^{-1}$) & ($\rm 10^{51}~erg$) &  ($\rm 10^{48}~erg~s^{-1}$) & ($\rm 10^{51}~erg$) \\
  \hline
    $z = $2.695& 18.09$\pm$1.53 & 3.18$\pm$0.88 & 48.48$\pm$0.73 &  48.79 $\pm$ 2.19 &  8.73 $\pm$ 0.40 
             & 1.08 $\pm$ 0.05 & 4.44 $\pm$ 0.23 \\
  \hline
    $z = $2.035 & 11.12$\pm$0.94 & 1.96$\pm$0.54 & 29.80$\pm$0.05 & 29.99 $\pm$ 1.35 &  6.53 $\pm$ 0.30  
              & 0.67 $\pm$ 0.03 & 3.32 $\pm$ 0.17 \\
  \hline
    $z=$ 1.695 & 8.00$\pm$0.68 & 1.41$\pm$0.39 & 21.43$\pm$0.32 & 21.56 $\pm$ 0.97 & 5.29 $\pm$ 0.24 
              &  0.48$\pm$ 0.02 &  2.69 $\pm$ 0.14 \\
  \hline
\end{tabular}
\end{center}}
\end{table*}

\begin{figure*}[htbp!]\label{fig:1}
\center
\includegraphics[angle=0,width=0.8\textwidth]{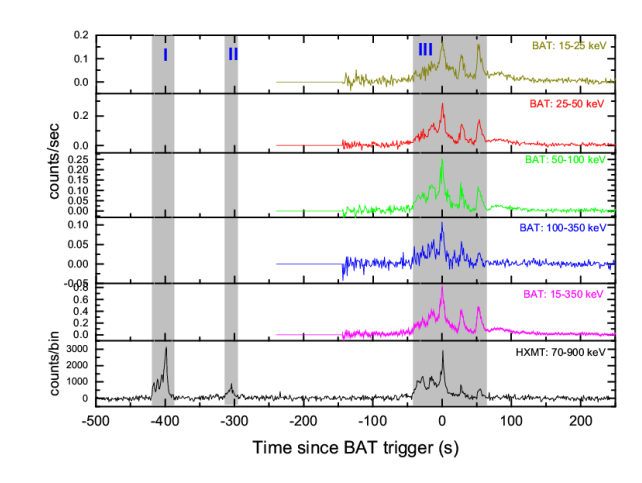}
\caption{Swift/BAT and HXMT light curves of GRB 240529A in different energy bands with a 1.0 second time-bin.}
\end{figure*}

\begin{figure*}[htbp!]\label{fig:2}
\center
\includegraphics[angle=0,width=0.5\textwidth]{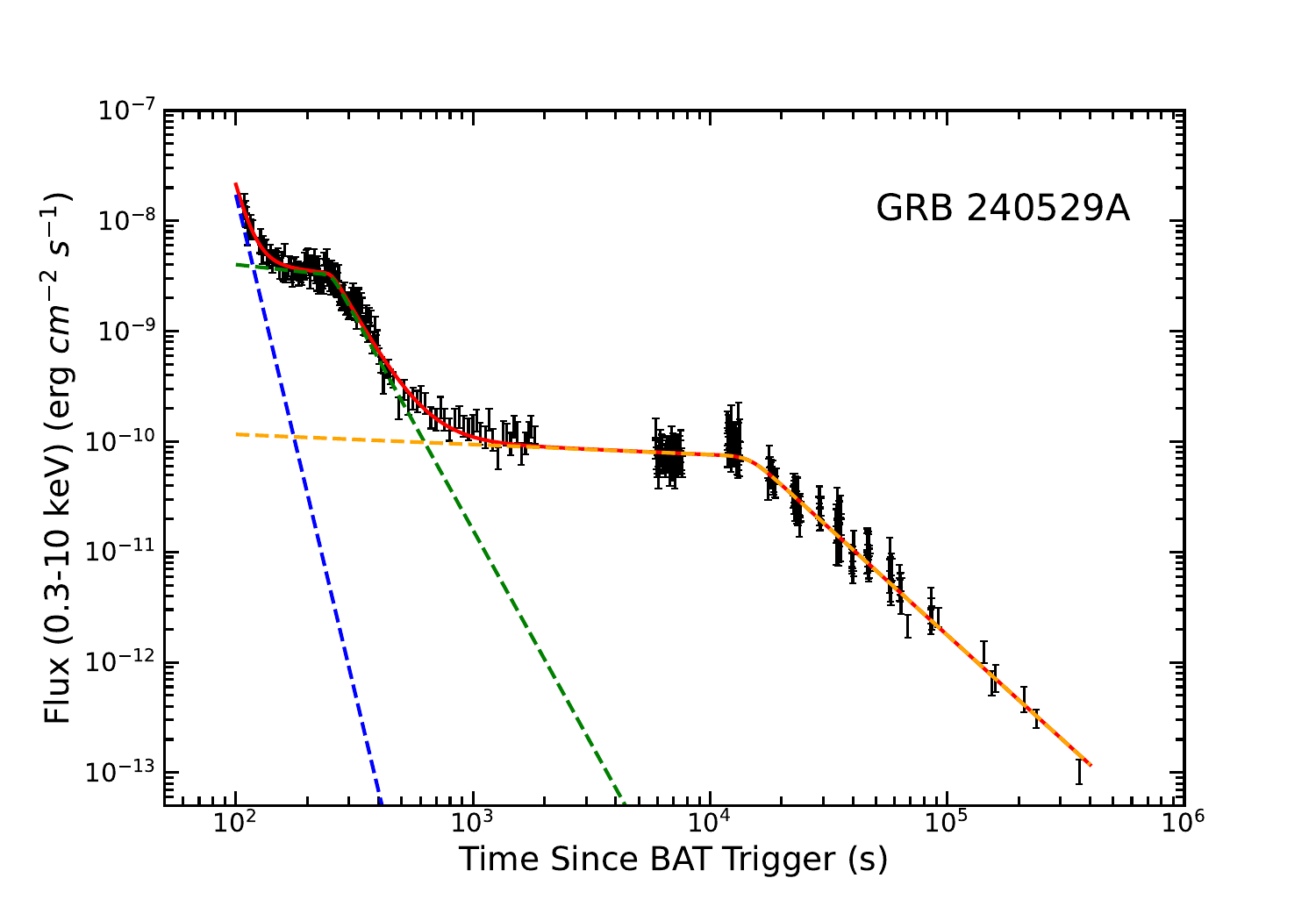}
\includegraphics[angle=0,width=0.4\textwidth]{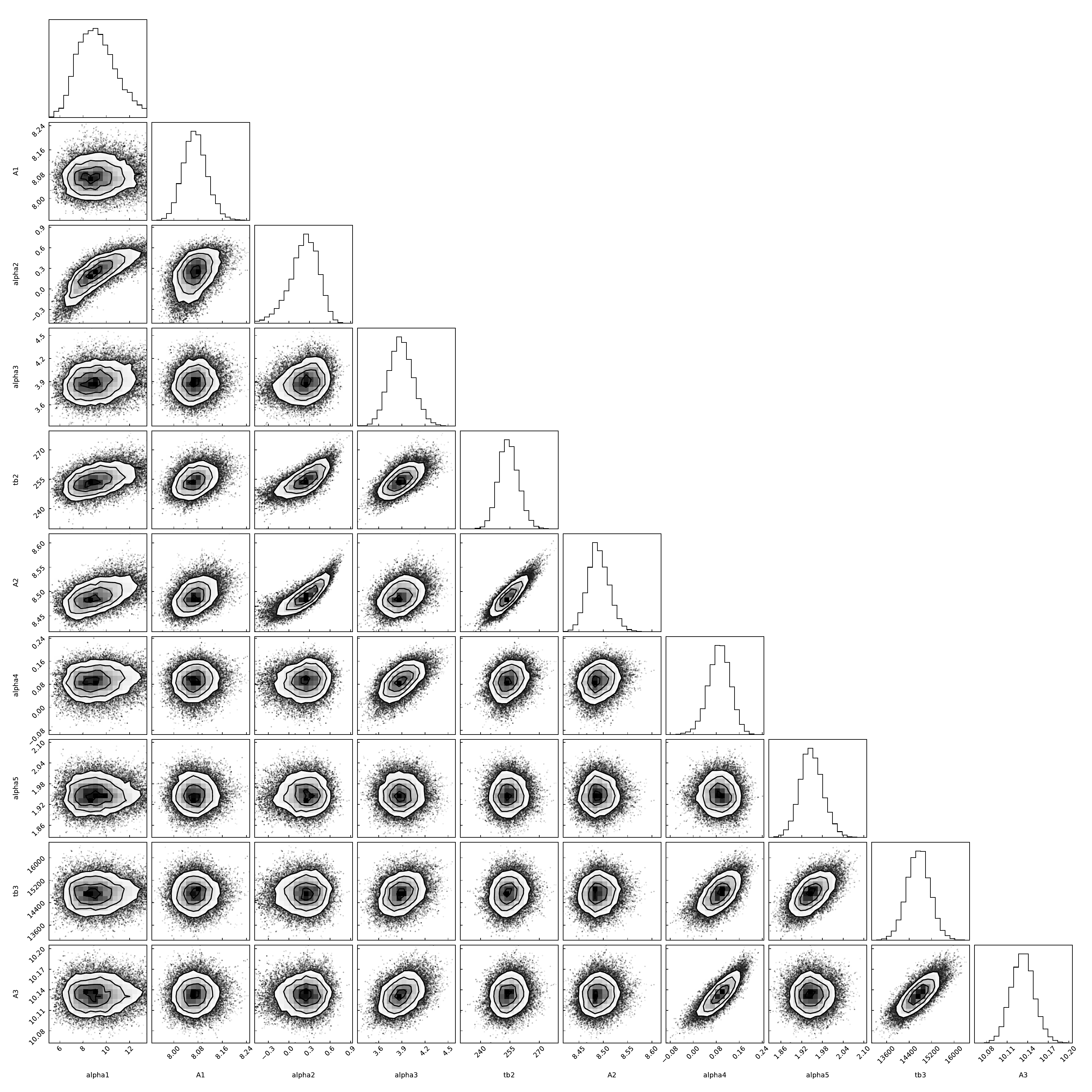}
\caption{Left: XRT (black) light curve of GRB 240529A. The red solid line represent the total component fitting line. The blue, green, and orange dashed lines are the best fitted lines for the power-law and two plateaus, respectively. Right: Corner plots and parameter constraints for the fitting parameters of the XRT light curve of GRB 240529A.}
\end{figure*}
\begin{figure*}[htbp!]\label{fig:3}
\center
\includegraphics[angle=0,width=0.8\textwidth]{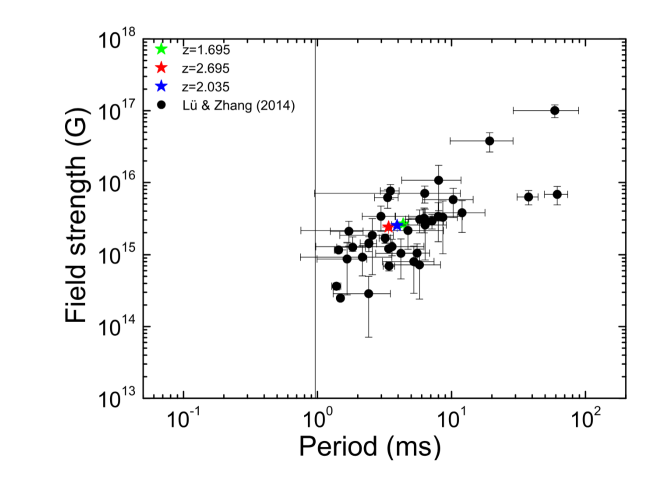}
\caption{Inferred strange star parameters of GRB 240529A, initial spin period $P_0$ vs. surface polar cap magnetic field strength $B_{\rm p}$ derived for different redshifts. The candidates of magnetar central engine are taken from \cite{2014ApJ...785...74L}. The vertical solid line is the breakup spin-period for a neutron star \citep{2004Sci...304..536L}.}
\end{figure*}

\begin{figure*}[htbp!]\label{fig:4}
\center
\includegraphics[angle=0,width=0.8\textwidth]{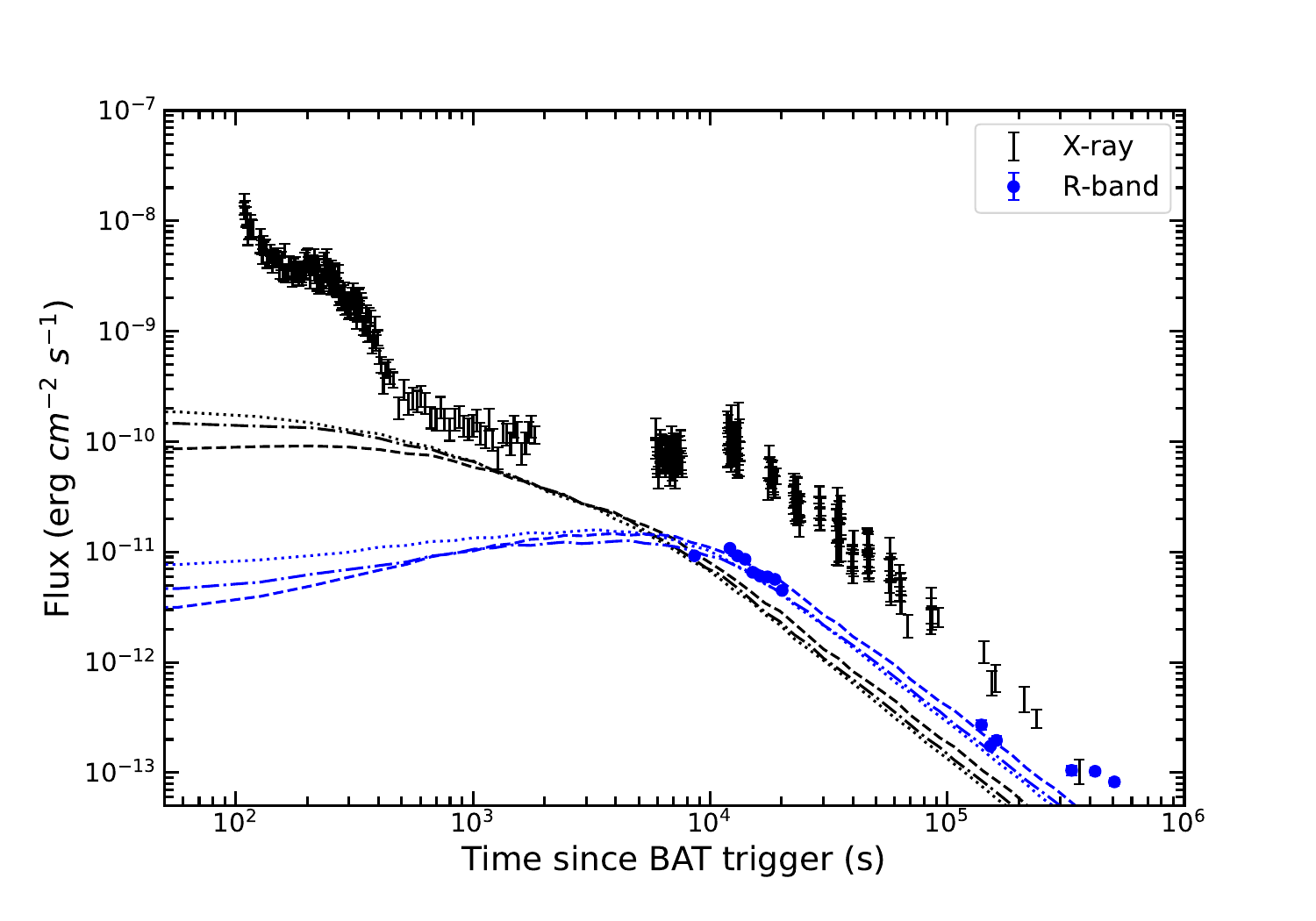}
\caption{The fitting results with external shock model in X-ray and optical bands of GRB 240529A. The standard afterglow model parameters show as: $\Gamma = 71$, $n=10$, $\epsilon_{\rm e}=\epsilon_{\rm B} = 0.08$ and $p = 2.8$. Here, the initial total kinetic energy $E_{\rm K,iso}$ is about $5\times 10^{52}$ erg, $2.5\times10^{52}$ erg, and $2.0\times10^{52}$ erg at redshift $z=2.695$ (dashed line), 2.035 (dash-dotted line), and 1.695 (dotted line), respectively.}
\end{figure*}


\end{document}